\documentstyle [prl,aps,epsfig]{revtex}

\newcommand{\be}{\begin{equation}}
\newcommand{\ee}{\end{equation}}
\newcommand{\bea}{\begin{eqnarray}}
\newcommand{\eea}{\end{eqnarray}}

\newcommand{\p}{\partial}
\newcommand{\s}{\sigma}

\newcommand{\rd}{\mbox{d}}
\newcommand{\ri}{\mbox{i}}
\newcommand{\re}{\mbox{e}}

\begin {document}
%\draft
\title{A Mechanism for  Ferrimagnetism and Incommensurability 
 in One-Dimensional Systems}
\author{A. M. Tsvelik}

%\vspace{0.5cm}

\address{ Department of Physics, University of Oxford, 1 Keble Road,
Oxford OX1 3NP, UK}

\maketitle
\begin{abstract}
%\par

 In this paper  I discuss a mechanism for ferrimagnetism in
 (1+1)-dimensions. The mechanism is related to a special 
 class of interactions described by operators 
with non-zero Lorentz spin. Such operators are present 
 in such problems as
the problem of tunneling between Luttinger liquids and the problem of
frustrated spin ladder. Exact solutions are presented   for  a
representative class of models possessing a  continuous isotopic 
symmetry. It is shown that  the  interactions   (i) dynamically 
generate static oscillations
with the  wave vector dependent on the coupling constant, (ii) give rise
to spontaneous breaking of this symmetry at $T = 0$ accompanied by
 generation of the magnetic moment and appearence of  
 gapless  modes  with a non-relativistic ({\it ferromagnetic})
dispersion $E \sim k^2$, (iii) generate massive (roton) modes. 
\end{abstract}

%\begin{multicols}{2}
%\narrowtext 
%\raggedcolumns

\section{Introduction}

 Though general stability of critical points is determined by scaling
 dimensions of the perturbing operators, to determine  
the ultimate destination
  of the renormalization group (RG) flow 
 is much more difficult problem. 
Availability of non-perturbative  
 methods makes this task easier in (1+1)- and two dimensions. 

 In
 (1+1)- or two dimensions 
  critical points possess conformal symmetry with the Lorentz (O(2))
 symmetry being its part. \footnote{Stricktly speaking, the Lorentz
 symmetry  is violated  in the models with  complex isotopic symmetry,
 where different sectors of the spectrum may have different
 velocities. However, this does not influence the argument made in the
 text.} If the relevant perturbation preserves the Lorentz symmetry,
 the most it can do is to open a gap in a part of the spectrum. That
 is exactly what happens in all known  solvable examples 
(for instance, in the sine-Gordon
 model). 

 It may happen however, that the perturbation violates the Lorentz
 symmetry. Such perturbations certainly make sense in condensed matter
 physics where preservation of the  Lorentz symmetry is not required. 
There are quite a few models of condensed matter physics which can be
 treated as critical models perturbed by relevant 
Lorentz-symmetry-breaking operators. These operators have non-zero
 Lorentz spin. Since in (1+1)- or two dimensions the Lorentz symmetry
 at criticality is extended to conformal symmetry, Lorentz spin is
 also called {\it conformal} spin.  
 At criticality where all two-point correlation functions follow power
 law behaviour, an operator $\hat O(z,\bar z)$ is  characterized 
by its  scaling dimension
 $d$ 
 and conformal (or Lorentz) spin $S$ defined as 
\bea
<\hat O(z_1,\bar z_1)\hat O(z_2,\bar z_2)> = Az_{12}^{-(d + S)}\bar
 z_{12}^{-(d - S)} \label{one}
\eea
 where $z,\bar z$ are holomorphic coordinates defined as 
$z = \tau + ix, \bar z = \tau - ix$ (in (1+1)-dimensions $\tau$ is
 Matsubara time). Since only  bosonic operators can appear 
 as perturbations, possible Lorentz spins are {\it integer}. 
 Furthermore, since in unitary theories conformal dimensions are
 positive ($d \pm S \geq 0$) and $d \leq 2$ for relevant operators,
 this leaves us with only one choice: $S = \pm 1$.

 The simplest   example of $S = \pm 1$ 
 operator  is a conserved
 charge. At criticality such perturbation generates 
 incommensurability. As an illustration  one can consider a change of the 
chemical potential
 for massless Dirac fermions:
\bea
A = \int \rd\tau \rd x[R^+\p R + L^+\bar\p L - \mu(R^+R + L^+L)]
\eea
This perturbation is removed by a simple transformation of the
 fields 
\bea
R \rightarrow \re^{\ri\mu x}R, ~~ L\rightarrow \re^{-\ri\mu x}L
\eea
and thus leads to a shift in the corresponding 
 Fermi momentum  and generates  static oscillations with the 
 new wave vector 
 whose  magnitude depends on the value of $\mu$.  The perturbed 
theory remains critical in the infrared (IR). 

 The situation becomes  less clear  
when the perturbation 
 is not  a constant of motion. Such non-trivial 
perturbation corresponds to  a non-holomorphic 
operator, that is, in the notations of Eq.(\ref{one}), an operator
 with  
 $d \neq \pm S$.

  A well studied example of $S= \pm 1$ perturbation is the 
 the so-called Z$_N$ 'chiral clock' model of statistical physics 
(see \cite{nijs} for review). In fact, this model is not too good for
 statistical physics purposes since its Boltzmann weights are not
 positively defined. However, the (1+1)-dimensional version of this
 model makes perfect sense and, as was demonstrated in \cite{nayak},
 describes a tunneling between two  spinless Luttinger liquids. 
It was established that for the chiral clock model 
 the relevant $S= \pm 1$  perturbations generate 
 incommensurability.

 An immediate generalization of the problem of tunneling includes
 spin. The  tunneling between two spin-1/2 Luttinger liquids is 
described by the
 Hamiltonian  
\bea
&&H = H_{LL}^{(1)} + H_{LL}^{(2)} + \int dx({\cal T}_{+} + {\cal T}_{-}), \\
&&{\cal T}_{+} = t (R^+_{1,\sigma}R_{2,\sigma} + H.c), ~~{\cal T}_{-} =
 t(L^+_{1,\sigma}L_{2,\sigma} + H.c) \label{tunn} 
\eea
where $R_{1,2}$ and $L_{1,2}$ are the right- and the left-moving fermions on
  chains 1 and 2; $H_{LL}$ describe interacting electrons on the
 individual chains. Due to the  intra-chain  interaction
 the fermionic operators acquire non-trivial scaling dimensions such
 that 
\bea
<{\cal T}_{\pm}(\tau,x){\cal T}_{\pm}(0,0)> \sim  \frac{1}{(\tau \pm
 ix)^2}\frac{1}{(\tau^2 + x^2)^{\theta}}
\eea
where $\theta$ depends on the interaction (here I do not discuss the
 effects related to difference between the charge $v_c$ and the spin $v_s$ 
velocities of
 the excitations and set $v_c = v_s = 1$). According to definition
 (\ref{one}) operators ${\cal T}_{\pm}$ carry conformal (Lorentz) 
spin $S = \pm 1$.

 Another problem much discussed in the literature is the problem of
 frustrated spin-1/2 two-leg  Heisenberg ladder (alias the zig-zag
 ladder). In the decoupled 
limit, two S=1/2 chains represent 
an $SU(2)_1 \times SU(2)_1$ WZNW theory. Each $SU(2)_1$ WZNW
model has its matrix field $\hat{g}_i (x)~(i=1,2)
= \epsilon_i (x) + \ri{\bf n}_i \cdot \vec{\s}$, (${\bf n}_{1,2}$ are staggered magnetizations of chains 1 and
 2 and $\epsilon_{1,2}(x)$ are  the staggered energy density operators). For
 the frustrated ladder the interchain interaction contains is
 dominated by the
 so-called  twist term  \cite{Ners98}. The most
general SU(2)-invariant form of this  term is 
\be
{\cal O}_{\rm twist} = A {\bf n}_1 \cdot \p_x {\bf n}_2
+ B \epsilon_1 \p_x \epsilon_2 + [1 \leftrightarrow 2]
\label{twist}
\ee
In the Heisenberg  zigzag ladder the bare value of $B$ is zero but it is
generated in the course of RG (due to the fusion of the $A$-operator
with scalar marginal ones). where the leading interchain interaction gives the following
 contribution to the action density  \cite{Ners98}:
 For a single chain the two-point
 function of the operators ${\bf n}, \epsilon$ decays as  
\bea
<{\bf n}(\tau,x) {\bf n}(0,0)> = <\epsilon(\tau,x) \epsilon(0,0)> 
\sim (x^2 + \tau^2)^{-1/2}
\eea
which means that 
\bea
<{\cal O}(\tau,x){\cal O}(0,0)> = \sim (x^2 + \tau^2)^{-1}[(\tau -
 \ri x)^{-2} + (\tau +
 \ri x)^{-2}] 
\eea
According to definition (\ref{one}),  this operator is
  a sum of two operators with $d = 2, S = \pm
 1$. 

 Using the procedure suggested in \cite{shelton}, one can reformulate
 the zig-zag ladder model as  a model  of four
 Majorana fermions with the Lagrangian density given by 
\bea
{\cal L} &=&{\cal L}_0 + {\cal L}_{twist}, \nonumber\\
{\cal L}_0 &=&\sum_{a =0}^3[\frac{1}{2}\chi_a(\partial_{\tau} 
- \ri\p_x)\chi_a +
 \frac{1}{2}\bar\chi_a(\partial_{\tau}
+ \ri\p_x)\bar\chi_a]\nonumber\\
{\cal L}_{twist} &= &  \{g_1\chi_1\chi_2\chi_3\bar\chi_0 +
 g_2[\chi_0\chi_1\chi_2\bar\chi_3 + \chi_3\chi_0\chi_1\bar\chi_2 +
 \chi_2\chi_3\chi_0\bar\chi_1] + [(\chi \leftrightarrow \bar\chi)]\}
 \label{zig} 
\eea
where $g_1, g_2$ are linear combinations of $A, B$. In the limit $A =
 B$ we get the perfect O(4) symmetry $g_1 = g_2$, with a suitable
 choice of $A,B$ one can get $g_2 =0$ which corresponds to the model I
 consider later in the paper.

 Operators with non-zero conformal spin may also appear in models
 describing  tunneling between edges of incompressible 
 Fractional Quantum Hall (FQH) states \cite{sondhi}.

 Since  operators with non-zero conformal spin break Lorentz symmetry,
 it lifts the restrictions on 
 the low energy spectrum. With the Lorentz symmetry
 being preserved one has two choices: either to have a linear spectrum
 or to have a spectral gap. Without Lorentz symmetry one can imagine a
 quadratic gapless spectrum which would be naturally associated with a 
 ferromagnetic Goldstone mode. Thus, if a model
 in question has  a continuous 
 isotopic symmetry, the presence of $S = \pm 1$ perturbation may lead
 to a spontaneously breaking of this symmetry  at $T =
 0$ with a formation of a net 'magnetic' moment and the
 above-mentioned quadratic spectrum.

 In all examples discussed above (except the one involving  the 
edge states) the perturbations  with  $S = \pm 1$ appear
together. This introduces an additional complification because 
 a fusion of two operators with opposite Lorentz  spins generates
 under RG a relevant operator with zero spin.  However, 
in a certain
 range of parameters of model (\ref{tunn}) this secondary  
flow does not catch up with  the flow
 of the original coupling constant (for model (\ref{twist} this is
 always the case)\cite{Kusm93},
\cite{Yakovenko92},\cite{Khv93} (see also
 \cite{Book}) where this topic is extensively discussed). 
This  gives room for  an energy
 scale on which the  
 effect of interaction between the  operators with $S = 1$ and $S = -1$
  can be neglected. On this energy scale one can drop one operator and
 consider a
 perturbation with one sign of conformal spin, 
for example $S = 1$ \cite{Ners00}. In FQH systems, where excitations
 are chiral, appearence of a single $S = \pm 1$ perturbation is more
 likely.

\section{A solvable model}

  In the subsequent sections I discuss a series of exactly solvable 
models which
 realize   the symmetry
 breaking scenario outlined above. These  models 
 are related to Wess-Zumino-Novikov-Witten (WZNW) model. The latter
model together with the related lattice models  has been used
extensively to address the problems in  magnetism (see, for example,  
\cite{faddeev}, \cite{devega}, \cite{pedro}, \cite{zvyagin}). 
Some of these lattice models 
 are  rather similar to the zig-zag spin ladder
 (\ref{zig}). 

 The particular model I discuss is  described by the following 
 action:
\bea
A = W_k^* + h_{\bar a}\int \rd^2x :J^a(x)\Phi_{a,\bar a}(x): \label{action}  
\eea
where $W_k^*$ is the action of the {\it critical} 
Wess-Zumino-Novikov-Witten (WZNW) model on
 the SU(2) group with level $k \geq 2$, $\Phi_{a,\bar a}$ are the 
primary field  from
 the adjoint representation, $J_a$ are the left 
Kac-Moody currents and $h_a$ is a
 constant vector. The restriction on the value of $k$ is related to
 the fact that for $k =1$ field $\Phi_{a\bar a}$ does not exist. 
The suggested solution   can be easily generalized for
 other symmetry groups and  coset models.

 To make the discussion self-contained, I recall several basic facts
 about WZNW models. A WZNW model describes  a matrix field
 $g(x)$ defined on a group $G$ whose dynamics is governed by  the action 
\bea
W(g) = \frac{1}{16\lambda
\pi}\int \rd^2x \mbox{Tr}(\p_{\mu}g^{-1}\p_{\mu}g) +  
k\Gamma[g] \label{wznw1}\\
\Gamma[g] = - \frac{\ri}{24\pi}\int_0^{\infty} \rd\xi\int \rd^2x
\epsilon^{\alpha\beta\gamma}\mbox{Tr}(g^{-1}\p_{\alpha}gg^{-1}\p_{\beta}g
g^{-1}\p_{\gamma}g) \nonumber %\label{gam}
\eea
The action contains two parameters - the coupling constant $\lambda$
 and integer number $k$. The model has a global $G_R\times G_L$ symmetry being invariant 
 under the transformations
\bea
g(x) \rightarrow V g(x)U
\eea
Hence  there are two conserved charges associated with left and
 right global shifts of the matrix field $g$. In the perturbed model
 (\ref{action}) the right symmetry is broken, but the left symmetry
 is not. As we shall see below, this symmetry is broken spontaneously 
 at zero temperature. 

 Model (\ref{wznw1}) has  a stable critical point; the critical
 value of the coupling constant is $\lambda^* =k^{-1}$. In what follows I
 shall distinguish between WZNW model as such and {\it critical} WZNW
 model described by action (\ref{wznw1}) with $\lambda =k^{-1}$.

 At the critical point WZNW models possesses a higher symmetry: 
 its right (left) {\it currents} become holomorphic (antiholomorphic):
\bea
 \bar J = \frac{k}{2\pi}g^{-1}\bar\p g, ~~\p \bar J = 0,\nonumber\\ 
 J = - \frac{k}{2\pi}g\p g^{-1}, ~~\bar\p J = 0\label{eqmotion}
\eea
(where $\p = \p_z, ~~\bar\p = \p_{\bar z}; z = \tau + i x, ~~ \bar 
z = \tau - i x$) and satisfy the Kac-Moody algebra:
\bea
[J^a(x),J^b(y)] = \frac{\ri k}{4\pi}\delta^{ab}\delta'(x - y) +
\ri f^{abc}J^c(y)\delta(x - y) \label{eq:Katz}
\eea 

 The critical  WZNW model can be conveniently described using  the
Hamiltonian formalism:
\[
\hat H = \frac{2\pi}{k + c_v}\sum_{a = 1}^{D}\int
\rd x[:J^a(x)J^a(x): + :\bar J^a(x)\bar J^a(x):] 
\]
where $c_v$ is the quadratic Casimir  in the adjoint
representation. For the SU(2) group $c_v = 2$.

 At criticality any  field   which is local in $g$ can be
 represented as a linear combination  of mutually local operators
 composing a basis in the operator space. This 
 basis of fields contains the primary fields $\Phi^{(j)}$ and their
 descendants. The primary fields transform under
 $G_R\times G_L$ as tensors belonging to irreducible representations
 of the group; the descendants are generated from the primary
 fields  by fusion with  the
 Kac-Moody currents. Primary fields from  higher
 representations  can be generated by fusion of the fields from the
 fundamental representation. However, this process terminates after
 certain  number of fusions  depending  on the value of $k$. 

 To make the discussion more concrete, I concentrate on the SU(2)
 group. In this case the primary fields are characterized by the value
 of spin $j = 1/2, 1, 3/2,...$. The field with the smallest spin $j =
 1/2$ is the $g$-matrix itself. A fusion of two $g$-operators
 generates a field with $j =1$. For the SU(2) group this
 representation is isomorphic to  the adjoint one. The maximal spin
 one can achieve by fusing $g$-matrices is equal to $k/2$. Therefore
 for $k =1$ model there is no primary field with $j =1$ and model 
(\ref{action}) is not well defined.

\section{Bethe ansatz solution}

 To solve   model (\ref{action}) I use 
the Bethe ansatz solution of  WZNW model \cite{PolWig83}. This
solution describes not just the critical point, but the entire
 RG flow towards it. As was shown in
\cite{Knizh84}, in the vicinity of the critical point the leading
irrelevant operator responsible for this  flow is  
\bea
V = \lambda :\bar J_a J_{\bar a}(x)\Phi_{a,\bar a}(x): \label{pert}  
\eea
In the presence of the right 
magnetic field ${\bf H}_R$ this operator generates perturbation
(\ref{action}). Indeed,  this field  is coupled
only to the currents of right chirality $\bar J_{\bar a}$. 
 In a finite magnetic
field these   currents acquire a finite expectation value  
\be
<\bar J_b> =  \delta^{ab} (H^{(R)}_a/2\pi)
\ee
and therefore in the leading order in $H$ the irrelevant operator  
(\ref{pert})
 is transformed into a relevant perturbation (\ref{action}) with 
\be 
h^a = \lambda (H^{(R)}_a/2\pi)
\ee

 Thermodynamic Bethe Ansatz (TBA) equations for SU(2) WZNW model of
 level $k$ in magnetic field have the following form \cite{PolWig83}:
\bea
\epsilon_n(v) &=& Ts*\ln[1 + \re^{\epsilon_{n + 1}(v)/T}][1 + 
\re^{\epsilon_{n - 1}(v)/T}] 
- m\delta_{n,0}\re^{\pi v/2} - m
\delta_{n,k}\re^{-\pi v/2}, \label{TBA}\\
&&\lim_{n \rightarrow - \infty} \frac{\epsilon_n}{n} = - H_R, ~~\lim_{n \rightarrow \infty} \frac{\epsilon_n}{n} =  H_L \nonumber  
\eea
where $ n =
 -\infty, ... \infty$. $H_R$ and $H_L$ are right and left 
 'magnetic' fields corresponding to the two conserved charges. 
In accordance with  the above discussion, I will keep only the leading
 contribution in $H_R = H$ and keep  $H_L$  infinidecimal.

 The free energy is given by  
\bea
F/L = 
-  mT\int \rd v(\re^{\pi v/2}\delta_{n,0} +
 \delta_{n,k}\re^{- \pi v/2})\ln[1 + \re^{\epsilon_{n}(v)/T}]
\eea
($L$ is the system's size) and 
\[
s*f(v) = \int_{-\infty}^{\infty} \rd u \frac{f(u)}{4\cosh[\pi(v - u)/2]}
\]

 We first study these equations at $T = 0$. In this case all
 $\epsilon_n$ with negative $n$ are of order of $nH$,  $\epsilon_0(v)$
 is positive at $v \rightarrow -\infty$ and 
changes its sign at $B \sim - \ln(m/H)$, $\epsilon_n$ with positive
 $n$ are also positive except of $\epsilon_k(v)$ which changes its sign
 from negative to positive at some value $Q > 0$. The ground state
 energy can be expressed in terms of $\epsilon_0$ and $\epsilon_k$
 which satisfy the following integral equations:
\bea
\int_Q^{\infty}\rd u D(v - u)\epsilon_k(u)
 = 
- m\re^{- \pi v/2} + \int_{- \infty}^{-B}\rd u K(v -
 u)\epsilon_0(u) \label{epk}\\
\int_{- \infty}^{-B}\rd u D(v - u)\epsilon_0(u) = 
 - m\re^{\pi v/2} +
 H + \int_Q^{\infty}\rd u K(v -
 u)\epsilon_k(u) \label{ep0}
\eea
where the Fourier transforms of the kernels are 
\bea
D(\omega) = \frac{\tanh\omega \re^{k|\omega|}}{2\sinh k\omega}, ~~
 K(\omega) = \frac{\tanh\omega}{2\sinh k\omega}
\eea
The limits $B$ and $Q$ are determined by the conditions
 $\epsilon_0(-B) = 0,\epsilon_k(Q) = 0$. 

 To determine the spectrum, we also need to know the distribution densities
$\rho_0, \rho_k$ which satisfy similar equations:
\bea
\int_Q^{\infty}\rd u D(v - u)\rho_k(u)
 = \frac{m}{4}\re^{- \pi v/2} + \int_{- \infty}^{-B}\rd u K(v -
 u)\rho_0(u) \label{rhok}\\
\int_{- \infty}^{-B}\rd u D(v - u)\rho_0(u) =   \frac{m}{4}\re^{\pi v/2} +
  \int_Q^{\infty}\rd u K(v -
 u)\rho_k(u) \label{rho0}
\eea 
These distribution densities determine the conserved charges (right
and left 'magnetic' moments):
\bea
Q_L = \frac{1}{2}\int_Q^{\infty} \rd u\rho_k(u), ~~ Q_R =
\frac{1}{2}\int_{-\infty}^{-B} \rd u\rho_0(u)
\eea

Since we are interested only in the leading asymptotics in $H$, we can
 neglect $\epsilon_k$ in Eq.(\ref{ep0}) (and respectively $\rho_k$ in
Eq.(\ref{rho0})) and solve these equations  as  Wiener-Hopf
 ones. Eqs.(\ref{epk}, \ref{rhok}) in this case also become  Wiener-Hopf
 equations. The case $k = 2$ is somewhat special and will be treated
separately. For $k > 2$ 
 the solutions at large $Q,B$ (small $H/m$) are  given by 
\bea
\epsilon_0(v) &=& \int_{-\infty}^{\infty}\frac{\rd\omega}{2\pi} \re^{
 \ri\omega(v + B)}\epsilon_0^{(+)}(\omega),~~  
\epsilon_0^{(+)}(\omega) =  \frac{\pi H}{\ri\omega(- \pi +
 2\ri\omega)G^{(+)}(\omega)G^{(+)}(0)}, \label{e0}\\
\epsilon_k(v) &=& \int_{-\infty}^{\infty}\frac{\rd\omega}{2\pi} \re^{-
 \ri\omega(v - Q)}\epsilon_k^{(+)}(\omega), ~~\epsilon_k^{(+)}(\omega)
 =  \frac{\pi (k - 2) m\re^{- \pi Q/2}}{(\pi -
 2\ri\omega)(\pi - \ri k\omega)G^{(+)}(\omega)G^{(-)}(-\ri\pi/2)}, \label{energy}\\
\rho_k^{(+)}(\omega) &=&\epsilon_k^{(+)}(\omega)\frac{(\pi - \ri
k\omega)}
{2\pi(k - 2)} \label{density}  
\eea
where 
\bea
&&m\exp(- \pi B/2) = H\frac{G^{(+)}(\ri\pi/2)}{G^{(+)}(0)},\\
&&\exp(- \pi Q/2) = \beta_k(H/m)^{\frac{(k + 2)}{(k - 2)}}, \\
&&\beta_k = 
\left\{\frac{k\tan(\pi/k)}{2\pi(k +
2)[G^{(+)}(\ri\pi/k)]^2}\right\}^{k/(k -
2)}\left[\frac{G^{(+)}(\ri\pi/2)}{G^{(+)}(0)}\right]^{\frac{(k +
2)}{(k - 2)}} \nonumber
\eea
and 
\bea
G^{(-)}(\omega) =  \left(\frac{\ri\omega k +0}{\pi
\re}\right)^{-\frac{\ri\omega k}{\pi}}\frac{\Gamma(1 + \frac{\ri
k\omega}{\pi})\Gamma(\frac{1}{2} + \frac{\ri \omega}{\pi})}
{\sqrt{2\pi k}\Gamma(1 + \frac{\ri \omega}{\pi})}
\eea
with $G^{(-)}(\omega) = G^{(+)}(-\omega)$ and $D(\omega) =
G^{(-)}(\omega)G^{(+)}(\omega)$.

 To obtain the modified TBA equations describing model (\ref{action}) I simply
 replace $\epsilon_0$ in Eq.(\ref{TBA}) by its zero temperature 
approximate value (\ref{e0}). The result is 
\bea
\epsilon_n(v) &=& Ts*\ln[1 + \re^{\epsilon_{n + 1}(v)/T}][1 + 
\re^{\epsilon_{n - 1}(v)/T}] 
 + \delta_{n,1}s*\epsilon_0(v) - m
\delta_{n,k}\re^{-\pi v/2}, ~~ (n = 1,...)\label{TBA1}\\
&&\lim_{n \rightarrow \infty} \frac{\epsilon_n}{n} =  H_L \nonumber  
\eea
The free energy is then given by  
\bea
F/L = 
-  mT\int \rd v \re^{\pi v/2}\ln[1 + \re^{\epsilon_{k}(v)/T}]
\eea

 Let us return to $T = 0$ solution. From  equations ( \ref{energy},
 \ref{density})    we can extract the following information. First
 of all, I  observe that $\epsilon_k$ approaches zero at two points:
 at $v = Q$ and $v \rightarrow \infty$. The latter point I identify
 with zero momentum; then point $v = Q$ corresponds to 
the wave vector
\bea
\delta P &=& 2\pi\int_Q^{\infty}\rd v \rho_k(v) =
 2\pi\rho_k^{(+)}(\omega = 0) 
=  m\gamma_k (H/m)^{\frac{(k + 2)}{(k - 2)}} 
\eea
where 
\[
\gamma_k = \frac{\beta_k}{G^{(+)}(\ri\pi/2)G^{(+)}(0)}
\]
This incommensurate wave vector scales with $H$  exactly
as one expects it to scale (that is $\delta P \sim H^{1/(2 - d)}$), 
taking into account that the scaling
dimension of the perturbation in Eq.(\ref{action}) is $d = 1 + 4/(k +
2)$. The excitation spectrum in  the vicinity of $v = Q$ is linear; the
velocity is given by 
\bea
V_R = \frac{1}{2\pi\rho_k(Q)}\frac{\p\epsilon_k(v)}{\p v}|_{v = Q} =
\frac{(k - 2)}{k}V_L \label{vel}
\eea
where $V_L$ is the velocity of the left-moving particles (I have been
working in the system of units  where $V_L = 1$). 

 The fact that
 $\epsilon_k$ approaches zero also at $v \rightarrow \infty$ 
 means that there are soft modes at zero momentum. Since according to 
 Eqs.(\ref{energy}, \ref{density})  $\epsilon_k(\omega), 
 \rho_k(\omega)$  behave as $|\omega|$ at small $\omega$, their real
 space asymptotics  at $v \rightarrow
 \infty$ are 
$\epsilon_k \sim \rho_k
 \sim v^{-2}$. Since the momentum  
 is given by 
\bea
P(v) = 2\pi\int^{\infty}_v\rd v \rho_k(v)
\eea
this means that the spectrum is quadratic. 
Using this formula and Eq.(\ref{density})  we find the dispersion law:
\bea
\epsilon(P) = \frac{(k - 2)}{k}\frac{P^2}{\delta P}, ~~(P << \delta P) \label{quadr}
\eea
The maximum of the energy is reached at $P \sim \delta P$ and is of
 order of $\delta P$.

For $k = 2$ the  formulae for the velocity and the dispersion law 
should be modified:
\bea
Q = 2\pi/g + \frac{1}{\pi}\ln (4g), ~~ g = [HG^{(+)}(\ri\pi/2)/m]^2\\
V_R/V_L = \frac{1}{\pi^2}g \sim H^2,\\
\delta P = \frac{m}{H[G^{(+)}(\ri\pi/2)]^2}\exp\{- [\pi
 m/HG^{(+)}(\ri\pi/2)]^2\}
\eea

 Taking $T \rightarrow 0$ limit in  Eqs.(\ref{TBA1}) I obtain  the following
 expressions for the energies:
\bea
&&\epsilon_n(\omega) = \nonumber\\
&&\frac{\sinh n\omega}{\sinh k\omega}\re^{\ri\omega
 Q}\epsilon_k^{(+)}(\omega) + \frac{\sinh(k - n)\omega}{\sinh k\omega}\re^{-\ri\omega
 B}\epsilon_0^{(+)}(-\omega), \label{ene}\\
 &&  n <k\nonumber\\
&&\epsilon_{n + k}(\omega) = \re^{-
 n|\omega|}\epsilon_k^{(+)}(\omega)
\re^{\ri\omega Q}
\eea
Let us study asymptotics of these energy functions. In real space we
 have 
\bea
\epsilon_{n + k}(v) = \frac{1}{\pi}\int_Q^{\infty}\rd u\frac{n}{(v -
 u)^2 + n^2}\epsilon_k^{(+)}(u)
\eea
From this expression it follows that 
 $\epsilon_{n + k}(v)$ decay as $v^{-2}$ on both infinities. On the
 other hand, $\epsilon_n(v)$ with $n < k$ decay as $v^{-2}$ only at
 $v>> Q$. At $v < Q$ they have roton-like minima. For $Q - v >> 1$ I
 obtain  from Eq.(\ref{ene}):
\bea
\epsilon_n(v) \approx k^{-1}\sin(\pi n/k)\left[\epsilon_k^{+}(\ri\pi/k)\re^{\pi(v
 - Q)/k} + \epsilon_0^{+}(\ri\pi/k)\re^{- \pi(v + B)/k}\right]
 \nonumber\\
= 2k^{-1}\sin(\pi
 n/k)[\epsilon_k^{+}(\ri\pi/k)\epsilon_0^{+}(\ri\pi/k)]^{1/2}\re^{-
 \pi(Q + B)/2k}\cosh[\pi(v - v_0)/k]
\eea
which corresponds to the relativistic spectrum with  spectral
 gaps for $k \neq 2$ ($n = 1, ... k -1$) given by
\bea
M_n = M_0\sin(\pi n/k), ~~ 
M_0 = \left[\frac{2(k - 2)\cot(\pi/k)}{\pi k(k + 2)}\right]^{1/2}\delta
 P \label{gaps}
%\frac{\beta_k}{G^{(+)}(\ri\pi/2)}\left[\frac{2(k - 2)}{\pi(k +
% 2)\sin(\pi/k)}\right]^{1/2}(H/m)^{\frac{(k + 2)}{(k - 2)}}
\eea
For $k =2$ the gap is exponential in $H^{-2}$. The energy minimum for
 $\epsilon_n$ with $n = 1,...k -1$ occurs  at 
\bea
v_0 = Q - \frac{k}{2\pi}\ln\left\{\frac{(k - 2)\tan(\pi/k)}{4\pi(k + 2)[G^{(+)}(\ri\pi/k)]^2}\right\}
\eea
 This
means two things. The first one is that  
the massive modes are centered at {\it incommensurate} wave
vector. The second is that in the rapidity space they are very close
to Q. 

\section{Bethe ansatz derivation of the low energy effective action}

 Thus we have the following regions  in rapidity space where low-energy
 excitations are located: 

(i) at $v >> Q$ there are gapless modes $\epsilon_n$ ($n = 1,...$) with
 the spectrum $\sim v^{-2}$;

(ii) at $v << Q$ there are gapless modes $\epsilon_n$ ($n = k+ 1,...$) with
 the spectrum $\sim v^{-2}$;

(iii) at $v = Q$ there is a gapless mode $\epsilon_k$ with the
spectrum $\sim (v - Q)$;

(iv) at $v \approx Q$ there are massive modes $\epsilon_n$ ($n =
1,...k - 1$). The massive modes can be treated as low energy
excitations only for $k >> 1$ when their masses are much smaller than
$\delta P$. 

 In the momentum space the spectrum  in  sectors (i),(ii) is
 quadratic $\epsilon(P) \sim P^2$ (see Figs. 1,2), the spectrum in sector (iii) is
 linear $\epsilon(P) \sim |P - P(Q)|$ and the spectrum in sector (iv)
 is massive relativistic (see Fig.3). 

\begin{figure}
\begin{center}
\epsfig{file=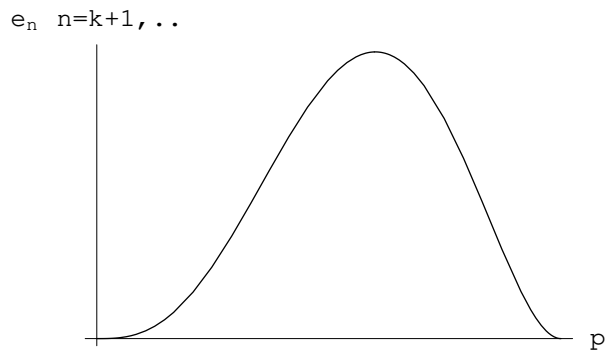,height=5cm}
\caption{Dispersion $\epsilon_n(P)$ ($n = k+ 1,...$)}
\end{center}
\end{figure}
\begin{figure}
\begin{center}
\epsfig{file=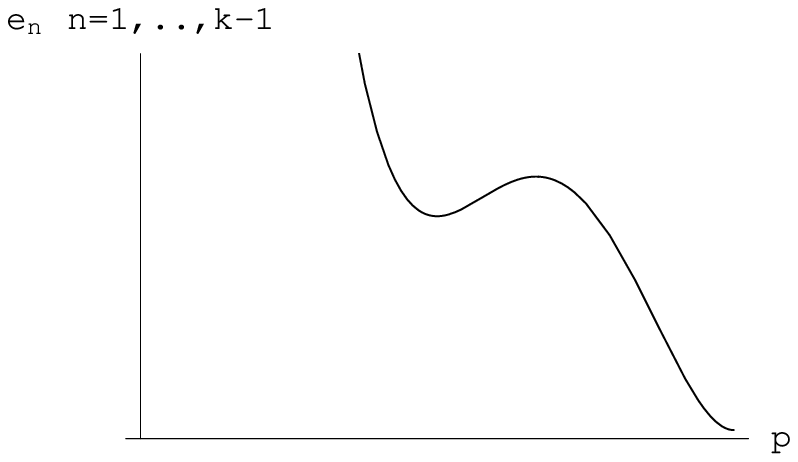,height=5cm}
\caption{Dispersion $\epsilon_n(P)$ ($n = 1, ... k-1$)}
\end{center}
\end{figure}

\begin{figure}
\begin{center}
\epsfig{file=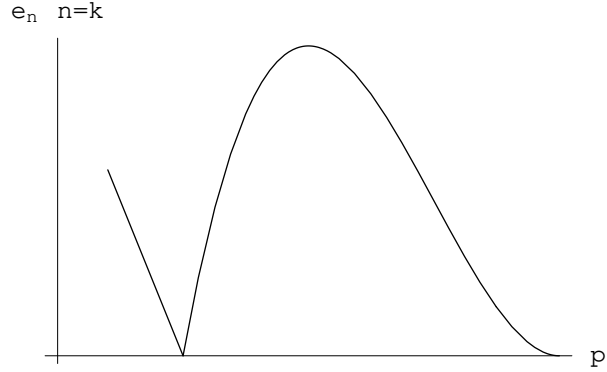,height=5cm}
\caption{Dispersion $\epsilon_k(P)$}
\end{center}
\end{figure}

  Let us first consider the case of moderate $k$ when one does not
 need to consider roton modes. Then the low energy sector is described
 by the  truely gapless modes (i - iii). Since  their positions in
 rapidity space  are well separated from each other 
 and the integration kernels in Eqs.(\ref{TBA1}) decay
 exponentially, one can derive separate sets of TBA equations
 for each mode. Comparing these equations with TBA
 equations for known integrable models one can deduce the effective
 action for the low energy sector.

 Let us derive  TBA equations for  sector (i). In this sector all
 energies are positive. Therefore it is convenient  to have TBA
 equations in such a form where the  integral kernels act on functions
 $\ln(1 +
 \re^{- \epsilon_n/T})$. Such  equations would provide a ready $T = 0$
 limit. To get such form of TBA. 
 The resulting
 equations read
\bea
T\ln[1 + \re^{\epsilon_n(v)/T}] - A_{nm}*T\ln[1 +
 \re^{-\epsilon_m(v)/T}] = A_{k,k}^{-1}*A_{n,k}*\epsilon_k^{(0)}(v) + nH_L, ~~ n= 1, ...,  ~~ v >> Q
\eea
where $\epsilon_k^{(0)}(v)$ is the solution at $T = 0$ and is given by
 Eq.(\ref{energy}). The kernels have the following standard Fourier
 transforms:
\bea
A_{nm}(\omega) = \coth|\omega|\left[\exp(- |n - m||\omega|) - \exp(-
 |n + m||\omega|)\right]
\eea
These equations are valid for temperatures $T << \delta P$ - the maximum
 value of $\epsilon_k^{(0)}(v)$. In this region 
These equations coincide with the  {\it right chiral} sector 
  of  the spin-(k/2) integrable
 {\it ferromagnet}. The latter ones  can be extracted from 
TBA equations   obtained in \cite{bab}. In a similar fashion I obtain
 TBA for sector (ii):
 \bea
T\ln[1 + \re^{\epsilon_{n + k}(v)/T}] - A_{nm}*T\ln[1 +
 \re^{-\epsilon_{m + k}(v)/T}] = a_n*\epsilon_k^{(0)} + nH_L, ~~ n= 1, ...,
 ~~ 
v <<  Q
\eea
These equations describe the  {\it left  chiral} sector of the
 spin-1/2  integrable
 {\it ferromagnet} where the dispersion law  is $\epsilon(k)
 \sim k^2\theta(k)$. Therefore the parity in the ferromagnetic sector
 is broken.

 Another soft mode in the right chiral sector has the linear 
  spectrum with velocity (\ref{vel}) and is centered at the the momentum
 $\delta P$. 
 Using the standard manipulations with  TBA equations one can find its
  contribution to the specific heat and establish  that this 
 mode carries central charge 1. Therefore it is described by the
 {\it non-chiral} Gaussian model. 

\section{Connection to the Zig-zag chain  at $k = 2$}
 
 For  $k = 2$ model (\ref{action}) simplifies considerably.  In this
 case the critical WZNW action is equivalent to the theory of three massless
 Majorana fermions $\chi_a, \bar\chi_a$ ($a = 1,2,3$)  and the
 adjoint operator and the currents are  given by \cite{ZamFat86}
\bea
\Phi_{ab} = \chi_a\bar\chi_b, ~~ J^a = \frac{\ri}{2}\epsilon^{abc}\chi_b\chi_c
\eea
Therefore for $k = 2$ one can rewrite 
the perturbed action
 (\ref{action}) solely in terms of Majorana fermions:
\bea
A =  \int\rd^2x\left[\frac{1}{2}\chi_a(\partial_{\tau} - \ri\p_x)\chi_a +
 \frac{1}{2}\bar\chi_a(\partial_{\tau} 
+  \ri\p_x)\bar\chi_a +  h_a\chi_1\chi_2\chi_3\bar\chi_a\right] \label{k2}
\eea
In this form the model closely resembles  the model for the zig-zag
 ladder (\ref{zig}) (see also  \cite{Ners00}) and att $g_2 =0$ it even
 coincides with  model (\ref{action})  exactly
 solvable by the Bethe ansatz.

{\bf Acknowledgements} I am grateful to Ph. Lecheminant for inspiration, to
 A. A. Nersesyan, D. Maslov 
 and F. H. L. Essler for illuminating discussions and
 interest to the work and E. Papa and S. Carr for help with preparation of the
 manuscript.

%\end{multicols}
\end{document}